\documentclass[prl,twocolumn,showpacs,epsf]{revtex4}
\usepackage{graphicx}
\begin{document}
\draft
\title
{   A Precision Measurement of the $^7$Be(p,$\gamma$)$^8$B Cross 
section with an Implanted $^7$Be Target}
\author{
L. T. Baby$^1$, C. Bordeanu$^1$\footnote{On leave from Horia Hulubei
 Institute, Bucharest, Romania.}, 
G. Goldring$^1$, M. Hass$^1$,  L. Weissman$^2$, V.N. Fedoseyev$^3$, 
U. K\"{o}ster$^3$, Y.Nir-El$^4$, G. Haquin$^4$, 
 H.W. G\"{a}ggeler$^5$, R. Weinreich$^5$ and the ISOLDE collaboration }
\affiliation{
{ 1.~~Department of Particle Physics, Weizmann Institute of Science, 
Rehovot, Israel}\\
{2.~~NSCL, Michigan State University, East Lansing, USA}\\
{ 3.~~ISOLDE, CERN, Geneva, Switzerland}\\
{ 4.~~Soreq Research Centre, Yavne, Israel}\\
{ 5.~~{ Paul Scherrer Institute,  Villigen, Switzerland}}\\}
\date{\today}

\begin{abstract}
The $^7$Be(p,$\gamma)^8$B reaction plays a central role in the evaluation of
 solar neutrino fluxes. We report on a new precision measurement of the
 cross section of this reaction, following our previous experiment with 
an implanted $^7$Be target, a raster-scanned beam and the elimination of the 
backscattering loss.  The new measurement incorporates a more abundant  $^7$Be
target and a  number of improvements in design and procedure. The point
 at E$_{\rm lab}$ = 991 keV was measured several times under
 varying experimental conditions, yielding a value of 
S$_{17}$(E$_{\rm c.m.}$ = 850 keV) = 24.0 $\pm$ 0.5 eV~b. 
Measurements were carried out at lower energies as well. Due to the precise
 knowledge of the implanted $^7$Be density profile it was possible to 
reconstitute both the off- and on- resonance parts of the cross section and
 to obtain from the entire set of
 measurements an extrapolated value of S$_{17}$(0) = 21.2 $\pm$ 0.7 eV~b.

\end{abstract}
\maketitle

The study of fusion reactions in the sun, relevant to the observed 
solar neutrino shortfall, has been the subject of intensive  research 
\cite{Adel,Bahc1}. Recently, this subject has acquired additional 
significance  with the new results of the Super-Kamiokande \cite{SK} and  
SNO \cite{sno} experiments. The  $^7$Be(p,$\gamma$)$^8$B
reaction and the accurate determination of the astrophysical S$_{17}$(0)
factor is of great importance to these and other astrophysical studies 
\cite{Fiorentini,Barger,Lopez} since  $^8$B 
is   the major source of the  high-energy solar neutrinos. In previous
publications \cite {Weiss,Hass} we have demonstrated a new
method for measuring  the cross section of the $^7$Be(p,$\gamma$)$^8$B
reaction by overcoming  several of the recognized potential
 systematic errors in earlier measurements (see, e.g., \cite{Adel}).
Our method involved a small diameter implanted  $^7$Be target
 from ISOLDE(CERN), incorporating  the elimination of back-scattering  
loss of $^8$B through the use of an implanted  target,  and a raster
 scanned beam over an area larger than the target spot, avoiding the  
difficulties  encountered with targets  of poorly known  areal distribution.
 Several   experiments \cite{Hammach,Jung,Stried}  have recently published 
S$_{17}$(0) values of (3-10)\% accuracy, two of those \cite{Jung,Stried} using
similar methods to \cite{Hass}. However, there still exist large, up to 20\% 
discrepancies among experimental results as well as the 
extracted S$_{17}$(0) values of these measurements. The present work 
has been undertaken in order to address these discrepancies and to 
provide a new, firm input for the determination of this cross section  by
 exploiting fully  the  advantages of the implanted target: full knowledge 
of the target composition and the $^7$Be profile, target robustness and the 
ability to produce a secondary target of reduced activity to improve the 
conditions for the $\gamma$ calibration of the target. Another feature of the 
present work is a thin $\alpha$ detector and a relatively large solid angle,
 providing clean $\alpha$ spectra.

The general scheme of the experiment follows that of our previous 
publications \cite{Weiss,Hass}. For the case of a 
homogeneous beam impinging  on a  target smaller than the beam,  the reaction 
yield Y,  in terms of  the cross section $\sigma$, is  given by: 
\begin{equation}
\rm Y = \sigma {\rm dn_b \over dS} \rm n_t
\end{equation}
where $\rm dn_b/dS$ is the beam density  and n$ \rm _t$
 is the total number of $^7$Be atoms in the target spot. The target was 
switched periodically between a position in the proton beam and a position in 
front of the $\beta$ delayed-$\alpha$ detector. The counting efficiency for 
the entire cycle is $\eta_{\rm cycle}$ = 0.390 $\pm$ 0.001. In terms of the 
experimental parameters, this can be written as: 
\begin{equation}
\sigma(\rm E_{\rm c.m.}) = {\rm N_\alpha\over n_t} \left( {A \over N _p} 
\right)
{1 \over \eta_{\rm Be}\cdot\eta_{\rm cycle}}
\end{equation} 
where N$_\alpha$ is the number of measured $\alpha$ particles,
  $N \rm _p/A$
is the integrated current density through a collimator hole of area A and 
 $\it \eta_{\rm Be}$ is the geometrical detection efficiency 
of $^8$Be, which
is twice the detection efficiency of $\alpha$'s: 
 $\eta_{Be}$ = ${2\cdot\Omega_\alpha}\over{4\pi}$.

\begin{figure}\begin{center}
\includegraphics[scale =.35]{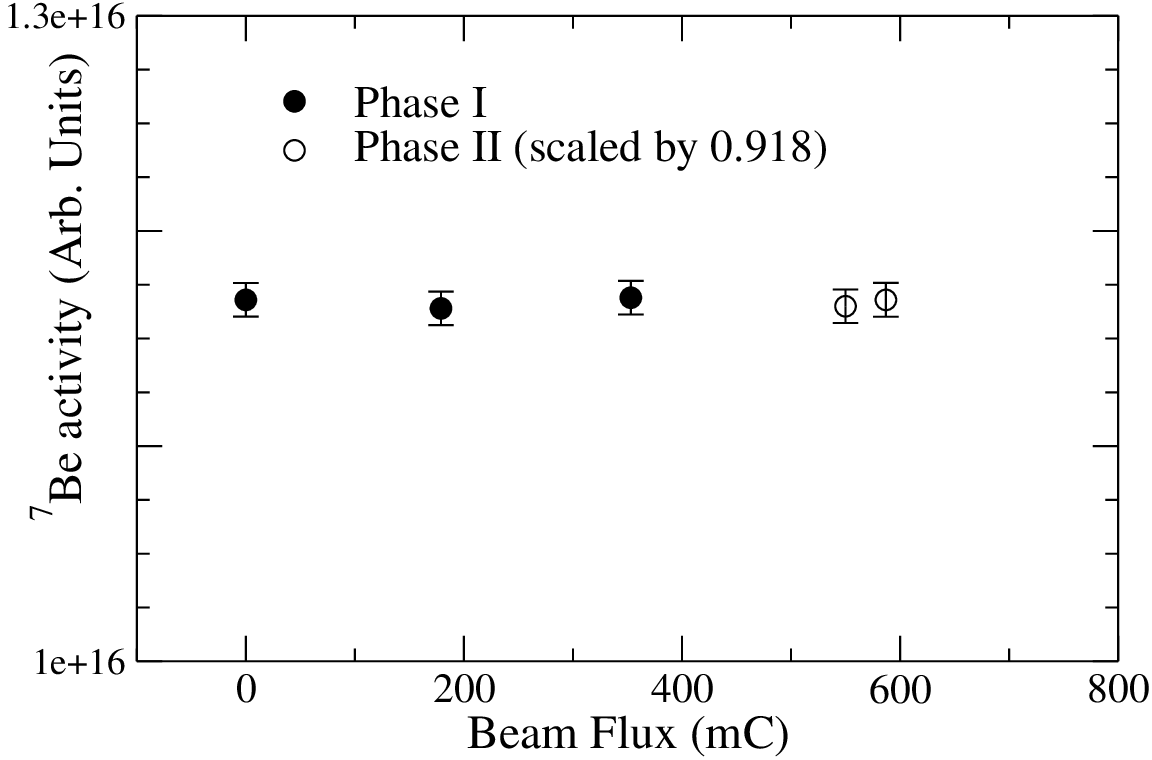}
\caption{The $^7$Be target activity monitored by a Ge detector as a function 
of the accumulated integral 
charge on target, normalized 
to the  total No. of 
$^7$Be just after implantation at ISOLDE and corrected for
the decay of $^7$Be. The  open circles  are scaled by a factor of 0.918 
(see text).}
\label{decay}
\end{center}
\end{figure} 
The proton beam from the Weizmann Institute 3 MV Van de Graaff accelerator
was scanned by an electronically controlled electrostatic raster scanner over a 
rectangular area in
 order to get a beam of uniform areal density \cite{Weiss1} passing  through a 
3 mm collimator placed  in front of the target. 
   A vacuum of 
$\approx$ 6$\cdot$10$^{-7}$ mbar was maintained in the chamber.
A liquid nitrogen cooled cryo-finger  was used  to reduce 
carbon build-up. The 2 mm diameter target spot was  aligned with a set of 
interchangeable
collimators downstream from the target, used to measure the beam density and 
homogeneity.  A 150 mm$^2$, 25 $\mu$m silicon
 surface barrier detector  recorded the 
$\beta$ delayed $\alpha$-particles from the decay of $^8$B. 
The target was mounted on an arm which was periodically 
rotated by a micro-step motor out of the beam  to face the detector
 at distances of 7-10 mm. The time
sequence of the cycle was similar to the one used in \cite{Hass}. The scanned 
beam density $\rm dn_b/dS$, typically of about 1-1.5 $\mu$A through a 2 mm
 aperture, was measured by integrating the  beam in an electron  suppressed
Faraday cup. The  current was digitized and recorded  in a 
gated scaler. Beam integration with and without suppression yielded results 
similar to a fraction of a percent. The beam homogeneity was checked by
measuring the $\alpha$ yield from the  $^7$Li(d,p)$^8$Li reaction  versus
integrated beam current for various downstream collimators. This procedure was 
repeated at various energies to obtain the corresponding scan voltage. The 
 $^7$Li(d,p)$^8$Li reaction was also used to determine the solid angle of the 
Si  detector by comparing the $\alpha$ yield of the 
$^7$Li(d,p)$^8$Li measurement (`close' geometry) with the yield in a `far'
geometry where the detector was collimated and far removed from
the target for  the solid angle to be  established directly. 
The proton energy from the Van de Graaff accelerator was calibrated 
with the help of the known resonances of the $^{27}$Al(p,$\gamma$)$^{28}$Si 
reaction at energies 991.2, 773.7, 632.6 and 504.9 keV.

\begin{figure}\begin{center}
\includegraphics[scale =.45]{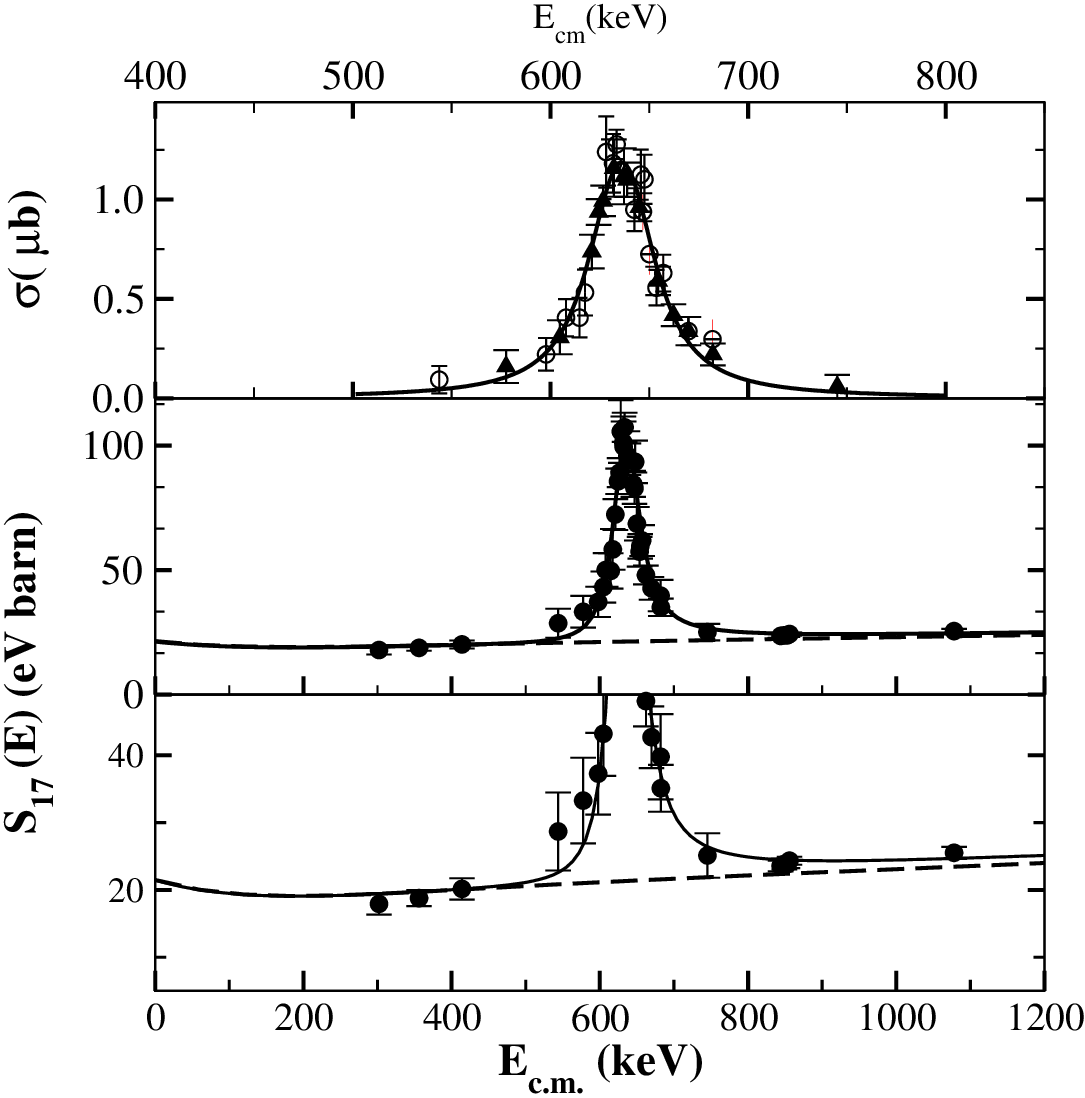}
\caption{Top:-The measured resonance at 633 keV. The  points are the measured
 cross section after subtraction of the non-resonant part. The energy
axis gives an expanded view of the resonance in comparison to the middle panel.
 The two symbols
 represent the resonance measurements carried out  at two different times. 
The continuous 
curve represents the convolution  of a   Breit-Wigner resonance with the 
simulated $^7$Be distribution in Cu.
Middle:- S$_{17}$(E)  at different energies.
The continuous line shows the  scaled function of Descouvemont and Baye [DB] 
  \cite{DB}  plus a 
Breit-Wigner resonance with an  energy dependent width.
 The dashed line shows the scaled DB model.
 Bottom:- An expanded view of the middle figure.}
\label{sfac}
\end{center}
\end{figure}

A high activity $^7$Be target was prepared at ISOLDE (CERN) in a  
manner similar to 
that  described in \cite{Hass} by direct implantation 
of $^7$Be at 60 keV in a copper substrate. The main new component in the  
present implantation  was the primary source of $^7$Be for  ISOLDE: 
a graphite target from the Paul Scherrer Institute (PSI) \cite{PSI}, 
used  routinely  at PSI for the production of $\pi$-mesons.
Large number of spallation  products  are accumulated in the target, including
$^7$Be. The graphite target was placed inside an ISOLDE ion-source 
canister and brought to ISOLDE for an off-beam implantation.  
The number of  $^7$Be atoms at
 production was  1.17$\cdot$10$^{16}$  on an implantation spot of 2 mm
 in diameter. The major characteristics of such a target- the deposition depth
of 1220 nm, the stability of the Cu-Be matrix and the negligible back-scattering
 loss are described in \cite{Hass}. In order to test the simulated  profile
 of the Be implants, we  measured the resonance curve in the 
$^7$Be(p,$\gamma)^8$B reaction. The position of the resonance peak indicated 
a proton energy of E$_{\rm lab}$ = 741 keV, consistent with the energy loss 
in 1220 $\AA$ of Cu, the mean deposition depth obtained from simulation. 
The results, summarized in Fig. \ref{sfac}, yield 
$\Gamma_{\rm c.m.}$ = 35 $\pm$ 3 keV for the resonance width, in agreement with 
the value 37 $\pm$ 5 keV of Ref. \cite{Filip}. The full width at half maximum 
of the $^7$Be simulated profile is 1200 $\AA$ and the second moment is 
2.5$\cdot$10$^5$ $\AA^2$.

The $\gamma$ activity  of the $^7$Be target  was too  intense to be assayed
in a standard  $\gamma$ calibration set-up due to the problems associated with
large dead times in the $\gamma$ counting. We therefore prepared a target
identical to that used in the experiment but $\approx$300 times weaker. An 
accurate measurement of the relative intensities of the 478 keV 
$\gamma$-rays from the decay of $^7$Be for the two targets was  
carried out at the low background 
$\gamma$ counting laboratory of  NRC-Soreq by placing both at the same
distance  from the Ge counter, yielding  a ratio of 317.8 $\pm$ 0.8. This
 ratio was determined several times with consistent results. The absolute 
intensity of the weak target was measured at NRC-Soreq and also at
Texas A$\&$M University. Both measurements followed calibration procedures 
involving up to thirteen high-precision standard sources of ten radio-nuclides.
The Texas A$\&$M procedure is discussed in detail in Ref. \cite{hardy}.
These two determinations yield a
(weak) target $^7$Be atom number of (2.667 $\pm$ 0.018)$\cdot$10$^{13}$ and 
(2.650 $\pm$ 0.018)$\cdot$10$^{13}$, 
 respectively. We  determine the number of $^7$Be  nuclei in the target
at a date corresponding to the end of implantation at ISOLDE to be  $n_t$ = 
 (1.168 $\pm$ 0.008)$\cdot$10$^{16}$. The 
branching ratio for $\gamma$ emission in the decay of $^7$Be was taken to be
 (10.52 $\pm$ 0.06)\% \cite{ndata} and the half-life  53.29 $\pm$ 0.07 days 
\cite{tuli}.  
The activity of the  Be target was monitored  throughout the  experiment 
 and revealed that  there was no significant loss of $^7$Be 
over the entire measurement (Fig.\ref{decay}). A conservative limit on the loss
is: $\Delta$($^7$Be)$<$ 1\%/C of integrated beam 
flux, except for a specific event in the last phase of the experiment
(see below).
The $^7$Li content of the target also exhibited a similar long term stability.

\begin{figure}\begin{center}
\includegraphics[scale =.44]{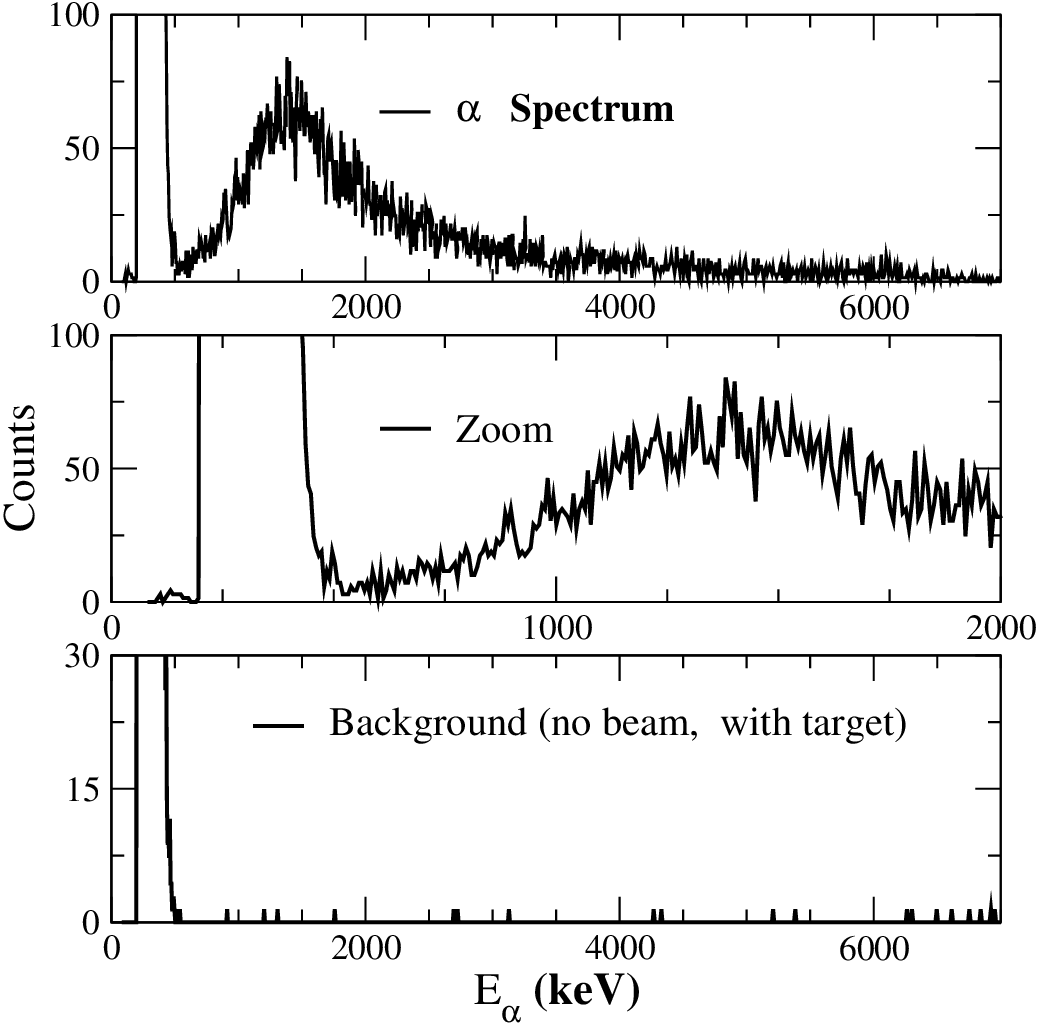}
\caption{Top: An $\alpha$ spectrum obtained at E$_{\rm lab}$ = 991 keV. 
Middle: An expanded view of the top panel. This spectrum 
was collected over a time of $\approx$ 80 hours.
Bottom: A background spectrum collected for $\approx$ 65 hours with the target
in the counting position and without beam.
The top and middle panels show the clear
 separation between the $\alpha$
and noise part of the spectrum. The noise cut-off is at E$_{\alpha}$ of 
$\approx$ 525 keV.} 
\label{alphasp}
\end{center}
\end{figure}

The 25 $\mu$m thick Si detector  provided a sufficient 
depletion layer to stop the $\alpha$ particles from the reaction but minimized 
the interaction with  $\gamma$ rays from the $^7$Be activity. 
Fig. \ref{alphasp} shows an $\alpha$ spectrum at E$_{\rm lab}$ = 991 keV.
The  $\alpha$ counts, N$_\alpha$ were obtained by integrating the spectrum of 
Fig. \ref{alphasp} in the region of interest.
The  number of $\alpha$ counts in the region of the fast rising noise peak
 is very small. For the data at E$_{\rm lab}$ = 991 keV  (table \ref{stable})
 this contribution has been 
determined as (0.3 $\pm$ 0.3)$\%$ to (0.8 $\pm$ 0.8)\% of the total. 

Background spectra were collected throughout the experiment
with the target in the counting position, both with beam-on and
beam-off and also with the target in the beam-through (current integration)    
position. The background consists of  four parts, 
 the electronic noise, pile up events from $\gamma$-activity, multiple
 scattering  of the beam and general background (no beam, no target).
  The electronic noise was 
appreciably reduced by water cooling the detector and it yielded no 
contribution to the recorded spectra. The contribution
to the background from  multiple scattering was minimized by tight shielding of
 both the detector and the proton beam path. It was determined to be negligibly
 small in a number of
 measurements, e.g., one in which the target was replaced by a blank target. 
The general  background was measured frequently; and the individual rates 
range from 0.4 to 0.6 $\pm$ 0.06/h. 
 The $\alpha$ counts from the reaction ranged from 100/h to 6/h.

\begin{table}
\begin{center}
\caption{The measured S$_{\rm 17}$(E) values along with  the details of the 
estimated error.
The  * indicates the set of measurements carried out at E$_{\rm lab}$ =
991 keV; the slightly different values of E$_{\rm c.m.}$ are due to
gradual carbon build-up, monitored by repeated measurements of the resonance.
Points indicated by **  were
 measured after the thermal episode as described in the text.
Columns 3-7 represent the contributions to the  error from counting
 statistics, background, beam energy, correction for $\alpha$ loss below the
region of interest, and a set of common errors:  the 
error on the number of $^7$Be atoms,  solid angle, timing efficiency, and
the area of the beam collimator.}

\begin{tabular}{cc|ccccc}
\hline 
E$_{\rm c.m.}$ &~~ S$_{17}$(E)~~&~Stat. &~ B.G~ &
Energy& $\alpha$-Cutoff & Common\\
 (keV)  & (eV~b) &  & & &  &\\
\hline \hline &\\
1078 	& 25.5 $\pm$ 0.9 	& 0.49 & 0.08 & 0.08 & 0.20 & 0.46\\ 
856$^*$ & 24.3 $\pm$ 0.6 	& 0.29 & 0.07 & 0.07 & 0.12 & 0.39\\
853$^*$ & 23.8 $\pm$ 0.6 	& 0.24 & 0.07 & 0.07 & 0.12 & 0.43\\
849$^*$ & 23.8 $\pm$ 0.8$^{**}$ & 0.44 & 0.07 & 0.10 & 0.19 & 0.49\\
844$^*$ & 23.6 $\pm$ 0.8$^{**}$ & 0.54 & 0.07 & 0.10 & 0.19 & 0.39\\
415     & 20.2 $\pm$ 1.6$^{**}$ & 1.36 & 0.41 & 0.36 & 0.1  & 0.42\\
356     & 18.8 $\pm$ 1.2$^{**}$ & 0.90 & 0.47 & 0.37 & 0.09 & 0.39\\
302     & 18.1 $\pm$ 2.0$^{**}$ & 1.50 & 0.54 & 0.80 & 0.09 & 0.40\\
\hline
\end{tabular}
\label{stable}
\end{center}
\end{table}
The cross section measurements were carried out at the energies of  
E$_{\rm c.m.}$ = 1078, 850, 415, 356 and 302 keV and also around the 633 keV 
resonance. The  carbon buildup on the target and the 
subsequent (small) energy degradation was monitored throughout the course of 
the experiment by repeating the resonance measurement  at regular intervals. 
The energy point at E$_{\rm c.m.}$ = 850 keV, corresponding 
to the $^{27}$Al(p,$\gamma)^{28}$Si resonance at E$_p$ = 991.2 keV, was 
measured several times in the course of this experiment  under varying
conditions of solid angle and target strength in order to serve as a benchmark
 (see Table \ref{stable}). 

In the last phase of the measurements the 
target was subjected to a thermal episode and was heated to a high temperature.
Following this event the $^7$Be content of the target dropped to 
0.918 $\pm$ 0.003 of the pre-event value. Before and after the event the values
were constant (taking into account the $^7$Be decay). A $^7$Be(p,$\gamma$)$^8$B 
resonance curve measurement revealed  that the second moment of the 
$^7$Be distribution  was broadened by the `event' by a factor of
 9. For most of our measurements the target is sufficiently thin 
so that  the  integration of the cross section over the  density distribution
can be inferred  to the requisite accuracy by correlating each measurements 
with the proton  energy at the  centroid of the $^7$Be distribution. 
This energy was 
derived  from the  measurement of the  peak of the 633 keV resonance.  Only 
at the 302 keV energy, there is an additional increase  in the cross section 
of 0.8\% because of the broadening.  

The measured S$_{17}$(E) values are shown in Fig. \ref{sfac} 
and in table \ref{stable}. The four measurements at E$_{\rm lab}$ = 991 keV, 
with the corresponding $\eta_{\rm Be}$ values of 0.1783, 0.2879, 0.2324,
0.1752, yield
 an average value  of S$_{17}$ = 24.0 $\pm$ 0.5  eV~b at the nominal energy of 
E$_{\rm c.m.}$ = 850 keV. This value provides a reference,  regardless of 
possible extrapolation  uncertainties, for comparison to other measurements.

The standard procedure at present is to employ   
 theoretical models for the extrapolation of S$_{17}$(E) to solar (``zero'') 
energy. This procedure is reasonable, as the nuclear physics processes involved
are rather simple and well understood. The practice of employing a 
generally adopted extrapolation model is supported by the observation that the 
disagreements among experiments are mostly in factors of proportionality in the
cross section  while the measured energy dependence is largely consistent. 
The present experiment utilizes  a 
thin target with a quite accurately known density profile. We have therefore
adopted the procedure of including all measurements, off and on the resonance,
in a fit with  the values of $\sigma_{\rm max}$ and
$\Gamma$ of the resonance and the overall normalization of the
generally used Descouvemont-Baye (DB) \cite{DB} theory as free parameters
 (Fig. \ref{sfac}).
We arrive in this way at a value of S$_{17}$(0) = 21.2 $\pm$ 0.7 eV~b.
We quote for completeness also the value derived from the low energy 
points (below the resonance) only as: S$_{17}$(0) = 20.8 $\pm$ 1.3 eV~b. 
The quoted S$_{17}$(0) values of all recent direct capture measurements
using a $^7$Be target are : 20.3 $\pm$ 1.2 eV$\cdot$barn,  
18.8 $\pm$ 1.7 eV$\cdot$barn,  22.3 $\pm$ 0.7 eV$\cdot$barn and  
 18.4 $\pm$ 1.6 eV$\cdot$barn (Refs. \cite{Hass,Hammach,Jung,Stried}, 
respectively).  
The mean of these values is:
S$_{17}$(0) = 21.1 $\pm$ 0.4 eV b with $\chi^2$/$\nu$ = 2.0, suggesting a
discrepancy. 
 If we omit the  value of 
ref. \cite{Jung} (which is being revised) from the list we get a mean value:
 S$_{17}$(0) = 20.5 $\pm$ 0.5 eV b with
$\chi^2$/$\nu$ = 1.2. If we add to this in quadrature an `error in theory'
 of ($\pm 0.5$),
as suggested in  ref. \cite{Jung}, we  get a consistent common value: 
S$_{17}$(0) = 20.5 $\pm$ 0.7 eV b.

A full  account of this work will be published elsewhere. 

We wish to thank the technical staff of ISOLDE(CERN), PSI,  and Y. Shachar and
the technical staff of the Accelerator Laboratory of the Weizmann Institute.
We acknowledge with gratitude the help of J.C. Hardy and V.E. Iacob with the
 intensity calibration of the $^7$Be target. We thank Prof. K.A. Snover
 for a fruitful exchange of information regarding the results of \cite{Jung} 
and work in progress. This work was supported in part 
by the Israel-Germany MINERVA Foundation.

\end{document}